\documentclass[prl,twocolumn,superscriptaddress,notitlepage,longbibliography]{revtex4-1}
\usepackage{amsmath}
\usepackage{amssymb}

\usepackage[unicode=true,colorlinks=true,citecolor=blue,urlcolor=blue]{hyperref}

\usepackage{bm}
\usepackage{epsfig}
\usepackage{multirow}

\usepackage[normalem]{ulem}

\renewcommand {\Re}{\mathop\mathrm{Re}\nolimits}

\renewcommand {\phi}{{\varphi}}
\newcommand {\rmi}{{\rm i}}

\newcommand {\e}{{\rm e}}

\begin{document}
\title{
{How single-photon nonlinearity is quenched with multiple quantum emitters:\\Quantum Zeno effect in collective interactions with  $\Lambda$-level atoms
}
}

\author{Alexander N. Poddubny}
\email{poddubny@weizmann.ac.il}
\affiliation{Department of  Physics of Complex Systems, Weizmann Institute of Science, Rehovot 7610001, Israel}

\author{Serge Rosenblum}
\affiliation{Department of   Condensed Matter Physics, Weizmann Institute of Science, Rehovot 7610001, Israel}
\author{Barak Dayan}
\affiliation{AMOS and Department of Chemical Physics, Weizmann Institute of Science, Rehovot 7610001, Israel}

\begin{abstract}
{
Single-photon nonlinearity, namely the change in the response of the system as the result of the interaction with a single photon, is generally considered an inherent property of a single quantum emitter.  Understanding the dependence of the nonlinearity on the number of emitters is important both fundamentally and practically, as strong light-matter coupling is more readily achieved through collective interactions than with a single emitter.
Here, we theoretically consider a system that explores the transition from a single to multiple emitters with a $\Lambda$-level scheme. We show that the single-photon nonlinearity indeed vanishes with the number of emitters. Interestingly, the mechanism behind this behavior is the quantum Zeno effect, manifested in the slowdown of
the photon-controlled dynamics. 
}
\end{abstract}
\maketitle 


Nonlinear behavior at the single-quantum level is  a  fundamental physical phenomenon that is also 
at the heart of quantum technology applications. In optics, the practical absence of photon-photon interactions at the single-photon level in a vacuum makes achieving single-photon nonlinearity  especially challenging.
Light-matter interaction can be enhanced in optical microresonators, enabling high single-atom cooperativity $C_1\gg1$  via Purcell enhancement \cite{Purcell,PhysRevLett.63.240}.  
Another approach relies on the collective response of multiple quantum emitters, either by using an ensemble of atoms~\cite{gorshkov2007,Hammerer2010,Corzo2019,Prasad2020,Bekenstein2020,Srakaew2023}, or  using a nonlinear crystal. However, the resulting nonlinearity in the latter case, for example of the $\chi^{(2)}$ or Kerr types, does not reach the level allowing deterministic single-photon nonlinear response such as those required for photon-atom logic gates \cite{Shapiro_2006}. { Interestingly, Rydberg excitations in atomic ensembles~\cite{Srakaew2023,Bekenstein2020} do not fully fall into the of category of collective response, since in order to achieve single-photon nonlinearity the Rydberg mechanism is harnessed to ascertain that only one atom is excited within the optical mode}. 
A deterministic single-photon response in systems with multiple quantum emitters still remains elusive. This is unfortunate, since obtaining high collective cooperativity $C_{N_{\rm at}}\equiv C_1 N_{\rm at}\gg 1$ with a larger number of  atoms is typically easier than reaching $C_1\gg 1$, which requires resonators with ultra-high quality factors and microscopic mode cross-sections. 

\begin{figure}[b]
\centering\includegraphics[width=0.48\textwidth]{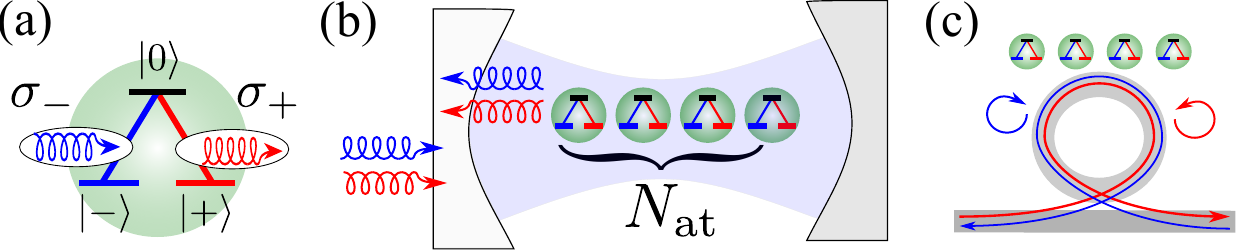}
\caption{  (a) Illustration of the two atomic transitions selectively coupled to $\sigma_+$- and $\sigma_-$ circularly polarized photons.  
(b) One-sided cavity with an array of $\Lambda$-atoms, coupled to right- and left-circularly polarized photons. (c) An equivalent realization with an array chirally coupled to a ring cavity. }\label{fig:1}
\end{figure}

Here we explore the vanishing of single-photon nonlinearity by considering an intriguing ``riddle" that involves multiple atoms in an optical cavity. 
Specifically, we consider atoms with a $\Lambda$-level configuration
with the two ground levels $|\pm\rangle$ and the excited state $|0\rangle$
(similarly to Kerr-type systems, the lack of ``memory'' in two level systems interacting with a single optical mode makes them insufficient for deterministic single-photon nonlinearity due to a time-bandwidth conflict \cite{Rosenblum_2011}).
The atoms are positioned in a single-sided cavity, namely a Fabry-P\'erot cavity with one perfect mirror and one mirror that serves as the input/output port. {Two modes of the cavity equally enhance} transitions corresponding to the two ``legs" of the $\Lambda$ system (here taken to be  $\sigma_+$ and $\sigma_-$ polarizations), as depicted in Fig.~\ref{fig:1}a.
A single $\Lambda$-atom in such a single-sided cavity enables the implementation of a photon-atom SWAP gate \cite{Pinotsi_2008,Koshino2010,Bechler2018} through a single-photon Raman interaction (SPRINT)\cite{Rosenblum_2017}. This means that a photonic qubit, encoded as a single photon in any superposition of the optical modes of the cavity, will be mapped to the atomic qubit, and vice versa --- the photon returning from the cavity will carry the qubit initially encoded in the two ground states of the atom $|\pm\rangle$. This behavior unambiguously corresponds to deterministic nonlinearity at the single-photon level; for example, it can extract the first photon in a multiphoton pulse to the other mode, while keeping all the following photons in the same mode~ \cite{Rosenblum2015}. We now wish to explore  how this nonlinearity changes with the number of atoms $N_{\rm at}$ while keeping the collective cooperativity constant: $C_1 N_{\rm at} = \textrm{const} \gg 1$.\\

\begin{figure}[t]
\centering\includegraphics[width=0.49\textwidth]{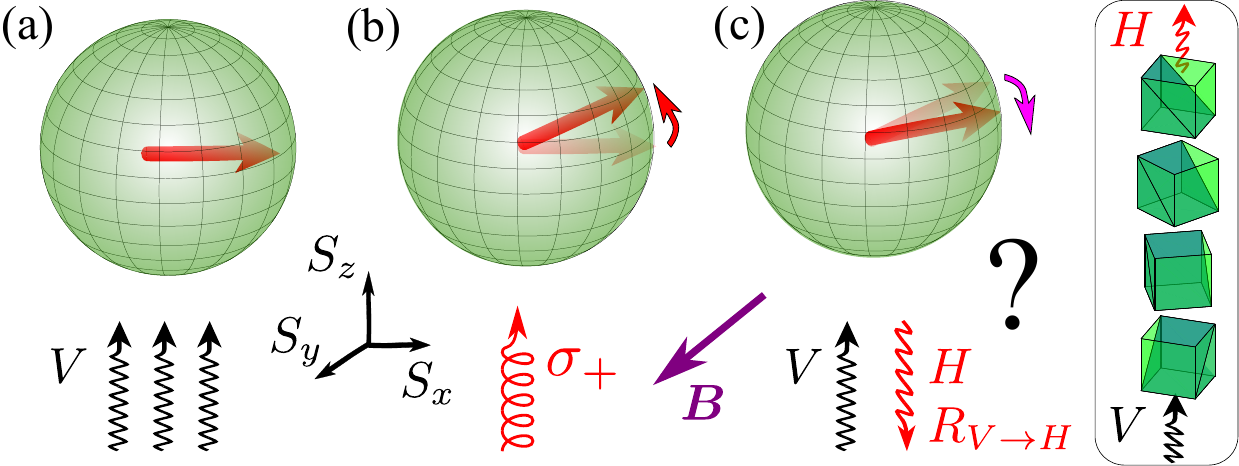}
\caption{Evolution of the collective spin of the $\Lambda$-atoms ensemble, driven by photons: (a) Spin is oriented along $x$ by a strong $V$-polarized excitation. (b) One $\sigma_{+}$ photon slightly rotates the spin towards  $z$. (c) The spin returns towards   $x$ after more $V$ photons are sent. Some of the sent photons are reflected with  $H$-polarization. External magnetic field $\bm B\parallel y$ can be also applied to rotate the spin.
{Inset in (c) illustrates  an analogy  with the quantum Zeno effect and nondemolition measurements: adiabatic photon polarization conversion by $N$ gradually rotating beamsplitters.}
} \label{fig:protocol}
\end{figure}

The riddle we consider is a three-step protocol, depicted in Fig.~\ref{fig:protocol}:\\
{(a)} By shining multiple $V$-polarized photons, all the atoms in the cavity are eventually pumped to the $H$ superposition (dark) state: 
\begin{equation}\label{eq:Sx}
\psi_{\rm in}=\frac{|+_1\rangle+|-_1\rangle}{\sqrt{2}}\otimes
\frac{|+_{2}\rangle+|-_2\rangle}{\sqrt{2}}\otimes\ldots\frac{|+_N\rangle+|-_N\rangle}{\sqrt{2}}\:
\end{equation}

\noindent {(b)} A single $\sigma_+$ polarized photon is incident on the cavity. The $V$ polarization component interacts  with a bare cavity, while the $H$ polarization interacts with a cavity  strongly coupled to multiple atoms. The $\pi$ phase shift associated with this condition flips the polarization of the photon to $\sigma_-$ as it reflects from the cavity. \\
{(c)} Clearly, the state of the atoms has now been perturbed, and cannot be assumed to remain the perfect dark state for $V$-polarized photons. We aim to verify this expectation by sending again many $V$-polarized photons, and asking: \textit{how many of these photons will return as $H$, and not stay $V$ as they would have, had we not sent the $\sigma_+$ photon in step (b)}? \\

We believe that this riddle is far from trivial. 
One could expect, for example, that since the dark state of the ensemble was modified by reflecting one 
photon, it will also need to reflect one 
photon in total before going back to its original state. This would amount to a strong single-photon nonlinearity. 
Another expectation is that this reflection process will be faster when the  ensemble has more atoms, as in the case of  Dicke superradiance~\cite{Dicke1954}.  We will show  that both these expectations are wrong.  
{The average total number of reflected $H$ photons $N_{\rm tot}^H$
is always below 0.5 (being the limit of a perfect SWAP gate with a single atom), and is suppressed by a factor of  $1/N_{\rm at}$ for a large number of atoms $1/N_{\rm at}\gg1$; moreover, the dynamics are slowed down and more $V$-photons are required to reach this value. }
This leads to the intriguing question: ``where  did the angular momentum  of the $\sigma_+$ photon (reflected as $\sigma_-$) go?''. 

{\it Quantum Zeno Effect}. We can gain some intuition  by considering another well-known scenario that involves repeated measurements: the Quantum Zeno Effect (QZE)~\cite{Misra1977,Koshino2005}. In the QZE, frequent quantum nondemolition (QND) measurements are applied at an axis that is identical to the initial state of the system. If the measurements are frequent enough so that the system can barely evolve or be perturbed between them, and assuming (as is usually the case) that for slight perturbations, the overlap with the original state drops quadratically, then the system is most likely to continuously collapse to the original state. This effect can be used either to ``freeze'' the evolution  (hence the name Zeno), or to adiabatically ``drag'' the state  by gradually changing the axis of the QND measurement. As an example,   consider a vertically polarized photon going through a series of $N$ cubic polarizing beam splitters, which gradually rotate by $90^\circ/N$ each until the last one is perpendicular to the first (inset in Fig.~\ref{fig:protocol}c). The probability that each beamsplitter will reflect the photon is proportional to $1/N^2$, and so the overall scattering probability goes down like $1/N$. The intriguing element here is that seemingly nothing is happening: the photon is never scattered, and yet its polarization is rotated. Here as well, we may ask ``where  did the momentum go?''. The answer is that the lack of scattering (quantum jumps) allows  coherent evolution to the new state, and the back-action is also transferred coherently to the environment, which is usually not part of the model. The force applied to the beam splitters accumulates coherently, and the momentum is transferred unnoticed through the mechanical holders of the beam splitters to the optical table and Earth. This is also the answer to our ``riddle'': the back-action (due to the momentum stored in the atomic ensemble in stage (b)) causes the $V$-polarized photons at stage (c) to become very slightly elliptical. {If we analyze these photons by a polarizing beam splitter, we get a similar situation: even though, for $N_{\rm at}\gg 1$, none of the photons is reflected to the $H$ port, 
a force is applied on that beam splitter until the momentum is coherently transferred from the atoms to Earth, and the atoms go back to their initial dark state (a).} With that intuition in mind, we  now present a more rigorous treatment.

{\it Model and theoretical framework.}
Besides answering the riddle above, our goal is to  introduce a rigorous theoretical approach for calculating photon scattering on $\Lambda$-atom ensembles. Despite the considerable recent progress,  current theoretical works still primarily focus on relatively simple states of the $\Lambda$-atom ensembles~\cite{Tsoi2009,Roy2011b,Martens_2013,Du2021,Zhong2023}, with the exception of only a few recent studies of  quantum squeezing \cite{Solomons2021,sundar2023squeezing}, superradiant bursts \cite{masson2023}, and photon cluster states for $N_{
\rm at}=1$~\cite{ilin2023manyphoton}.  The general problem of multiple photon scattering on $\Lambda$-atom ensembles remains unsolved to the best of our knowledge. 
 
We assume that the linewidth of the atoms is much smaller than the cavity linewidth. In this case, the Markovian approximation is valid and the  photonic degrees of freedom can be traced out~\cite{Caneva2015}. We  describe the ensemble by  the following effective Hamiltonian
$\hat H={\omega_0\sum_{n=1}^{N_{\rm at}}|0,n \rangle \langle 0,n|+}\sum_{\nu=\pm}\,\sum_{m,n=1}^{N_{\rm at}}\sigma_{\nu,n}^{\dag}\sigma_{\nu,m}^{\vphantom{\dag}} D_{nm}\:.$
Here, $\omega_0$ is the resonance frequency of the atomic transitions, and $\sigma_{\nu,n}^{\dag}$ are the corresponding raising operators. Since each {of the $N_{\rm at}$ atoms} has two ground states $|\pm ,n\rangle$ and a single excited state $|0,n \rangle$, the raising operators have the following nonzero matrix elements $\langle 0,n|\sigma_{+,n}^{\dag}|+,n\rangle=\langle 0,n|\sigma_{-,n}^{\dag}|-,n\rangle=1$. The matrix $D_{nm}$ is proportional to the electromagnetic field  Green function that describes a photon emitted at atom $m$ and reabsorbed at atom $n$.  
As shown in the Supplementary Materials, for a one-sided cavity with the interatomic distances negligible compared to the photon wavelength, one has 
$D_{nm}=-\rmi \gamma_{\rm 1D}/2$ and the Hamiltonian reduces to
\begin{equation} \label{eq:H1}
\hat H={\omega_0}\sum\limits_{n=1}^{N_{\rm at}}|0,n \rangle \langle 0,n|-\rmi\left(\frac{\gamma_{\rm 1D}}{2}+\gamma\right)\sum\limits_{\nu=\pm}\sigma_\nu^{{\rm tot},\dag}\sigma_\nu^{{\rm tot}}\:.
\end{equation}
Here, just like for two-level atom ensembles described by the celebrated Dicke model~\cite{Dicke1954,gross1982}, the collective interaction with photons involves only the total spin operators $\sigma_\nu^{{\rm tot}}=\sum_{n=1}^{N_{\rm at}}\sigma_{\nu,n}$, with $\nu=+$ or $-$.
{The phenomenological decay rate $\gamma$ accounts for emission into nonresonant photon modes and other decay processes.}
 Our results apply not only to the one-sided cavity in Fig.~\ref{fig:1}(b) but  also to  a ring cavity coupled to a waveguide, illustrated in Fig.~\ref{fig:1}(c), and experimentally realized in Refs.~\cite{Dayan2008,Rosenblum2015,Bechler2018}.

We are interested in the weak excitation limit, when the array is illuminated by photons one by one, so that the array never becomes doubly excited. The optical transitions will then occur only between the $2^N$ ground states $|\pm ,n\rangle$ and the $N2^{N-1}$ single-excited  states where exactly one atom is excited to the state $| 0,n\rangle$, that we will label $|\psi_1^{(\mu)}\rangle$.
We also introduce the Green function 
\begin{equation} 
G(\omega)=\rmi \gamma_{\rm 1D}\sum\limits_{\mu\mu'} |\psi_1^{(\mu)}\rangle \left[\frac1{H_1-\hat 1\omega}\right]_{\mu\mu'}\langle\psi_1^{(\mu')}|,
\end{equation}
where 
$[H_1]_{\mu\mu'}\equiv \langle\psi_1^{(\mu)}| H  |\psi_1^{(\mu')}\rangle\:.
$ 
The calculation, detailed in the Supplementary Materials, provides the following expression for the  state of the atomic ensemble after the photon scattering event:
\begin{equation}\label{eq:final}
\psi_{\rm scat,\nu\rightarrow\nu'}=[\delta_{\nu,\nu'}+\sigma^{\rm tot}_{\nu'} G(\omega) \sigma^{\rm tot,\dag}_{\nu}]   \psi_{\rm ground}\:,
\end{equation}
where  $\psi_{\rm ground}$ is the state before the scattering and $\nu,\nu'$ are incident and scattered photon polarizations.   The first term in Eq.~\eqref{eq:final} describes a photon reflected directly from the left cavity mirror, without interacting with atoms, while the second term involves  photon interaction with the single-excited states. 
The corresponding photon reflection coefficient is given by the expectation value $R_{\nu\rightarrow\nu'}\equiv\langle \psi_{\rm scat,\nu\rightarrow\nu'}|\psi_{\rm scat,\nu\rightarrow\nu'}\rangle$. The atomic state transformation  Eq.~\eqref{eq:final} is unitary for $\gamma=0$, that is $\sum_{\nu'}R_{\nu\rightarrow\nu'}=1$.

We  apply Eq.~\eqref{eq:final} iteratively to describe consecutive interaction with multiple photons. At each iteration,  we account for all the possible polarizations of scattered photons $\nu'$. This yields a tree-like record of  scattering events $\nu\to\nu'\to\nu''\to\ldots$ with different atomic ensemble wavefunctions and photon reflection coefficients assigned to each node of the tree.
The expectation value of the total number of reflected photons  $N_{\rm tot}^H$ is obtained as a sum of reflection coefficients over all scattering configurations. For example, if the incident photons  have $V$ polarization and we are counting reflected photons in $H$ polarization, for the first incident photon, $N_V=1$, the reflection coefficient is  $R_{V\to H}^{(1)}\equiv\langle \psi_{{\rm scat},V\to H}|\psi_{{\rm scat},V\to H}\rangle$.
Before the second $V$-polarized photon has arrived, the ensemble is in the state 
 $\psi_{{\rm scat},V\to H }$ with probability $R_{V\to H}^{(1)}$ and in the state $\psi_{{\rm scat},V\to V}$ with probability $R_{V\to V}^{(1)}=1-R_{V\to H}^{(1)}$. These two states serve as two possible  ground states $\psi^{(a)}_{\rm ground}$ and
 $\psi^{(b)}_{\rm ground}$. For each of the ground states, we can use   Eq.~\eqref{eq:final} to find the probability of detecting a second reflected photon with $H$ polarization, that is $R_{V\to H}^{(a,b)}$.
  The  reflection coefficient  for the second $V$ photon, averaged over the initial state, is $R_{V\to H}^{(2)}$=
 $R_{V\to H}^{(1)}R_{V\to H}^{(a)}+R_{V\to V}^{(1)}R_{V\to H}^{(b)}$.

{\it Spin  dynamics and QZE.}
By considering multiple $V$-polarized photons,  we have verified that the state Eq.~\eqref{eq:Sx} is  indeed the equilibrium state of ensemble after step (a) of the ``riddle'' protocol in Fig.~\ref{fig:protocol}. This can be also understood intuitively by noticing  that this  state does not interact with   $V$-photons due to the parity selection rules.

Next, we discuss the effect of the  protocol steps in Fig.~\ref{fig:protocol}(b,c) on  the collective atomic spin.  We introduce  a (pseudo)spin-1/2 $\bm s^{(j)}$ operator
{acting in the space spanned by the two ground states of the atom, e.g.
$s_z^{(j)}|\pm\rangle=\pm |\pm\rangle/2$.} Because of the symmetry of the problem, incoming photons are coupled only to the collective spin of the array $\bm S=\sum_{j=1}^{N_{\rm at}} \bm s^{(j)}$. After the initialization step the collective spin is oriented along the $x$ direction and is equal to $N_\textrm{at}/2$, see   Fig.~\ref{fig:protocol}(a). 
Scattering of the circularly polarized photon transfers the angular momentum to the ensemble, and the collective spin is rotated in the $x-z$ plane towards $z$ (Fig.~\ref{fig:protocol}b).  Indeed,
as follows from Eq.~\eqref{eq:final}, this scattering process is described by the operator $\sigma^{\rm tot}_{-} G(\omega) \sigma^{\rm tot,\dag}_{+}$.  As shown in the Supplementary Materials, 
$\sigma^{\rm tot}_{-} G(\omega) \sigma^{\rm tot,\dag}_{+}=\rmi \alpha(\omega) (S_x-\rmi S_y)\:,$
where
\begin{equation}\label{eq:alpha}
 \alpha(\omega) =-\frac{\gamma_{\rm 1D}}{\omega-\omega_0+\frac{\rmi}{2}\gamma_{\rm 1D}(N_{\rm at}+1)+{\rmi \gamma}}\:
\end{equation}
is the effective polarizability. The imaginary term $\frac{\rmi}{2}\gamma_{\rm 1D}(N_{\rm at}+1)$ in the denominator of  Eq.~\eqref{eq:alpha} increases linearly with  $N_{\rm at}$,  reflecting collective enhancement of the atom-photon coupling. 
{The ratio between the radiative and nonradiative decay rates in the denominator gives the collective cooperativity
$C_{N_{\rm at}}=\gamma_{\rm 1D}(N_{\rm at}+1)/(2\gamma)$.} 
The spin operator $S_x-\rmi S_y$ describes the increase of the absolute value of the {$z$ projection of the} collective pseudospin by 1 during the scattering of the first circularly polarized photon. We expect that at the next step, Fig.~\ref{fig:protocol}(c),  after the interaction with $V$-polarized photons the collective spin will {rotate  in the $x-z$ plane back }to its original direction along $x$, described by Eq.~\eqref{eq:Sx}, but some light will be  reflected in $H$-polarization during this relaxation stage. This directly follows from the matrix elements of atom-photon interaction. The relevant operators describing spin evolution are  
$\sigma^{\rm tot}_{V} G \sigma^{\rm tot,\dag}_{V}=\rmi\alpha (\tfrac{N_{\rm at}}{2}-S_x)$, and
$\sigma^{\rm tot}_{H} G \sigma^{\rm tot,\dag}_{V}=\rmi\alpha (S_z-\rmi S_y)$\:.
Hence, application of a large number of $V$ photons should drive the spin to the eigenstate of $S_x$ operator. {After the 
first $\sigma_+$ photon is scattered we obtain  $ \langle S_z\rangle \ne 0 $, $\langle S_y\rangle=0$,}
and as long as $\langle S_z  \rangle$ remains nonzero the 
expectation value of $\sigma^{\rm tot}_{H} G \sigma^{\rm tot,\dag}_{V}$ will also be nonzero and the array will be able to
reflect photons in $H$ polarization. In particular, the reflection coefficient of the first $V$ photon is given by the matrix element of 
$\sigma^{\rm tot}_{H} G \sigma^{\rm tot,\dag}_{V}$
between the states with $S_z=0$ and $S_z=1$, that is
$R_{V\to H}^{(1)}=|\alpha|^2$.
The final state, after a large number $N_V$ of incident $V$ photons,  will be the state with $\bm S\parallel x$, $S_x=N_{\rm at}/2$ and $S_z=0$, so that $V\to H$ photon scattering process will be no longer possible. It is this  pinning of the spin to the $x$ axis by subsequent photon scattering events that we interpret as a quantum Zeno effect~\cite{Misra1977,Kofman2000,Koshino2005,Leppenen2022,Gershon2022}.
Indeed, the spin dynamics is slowed down by  the observer trying to optically probe the spin state, which is the essence of QZE.

\begin{figure}[t]
    \centering\includegraphics[width=0.48\textwidth]{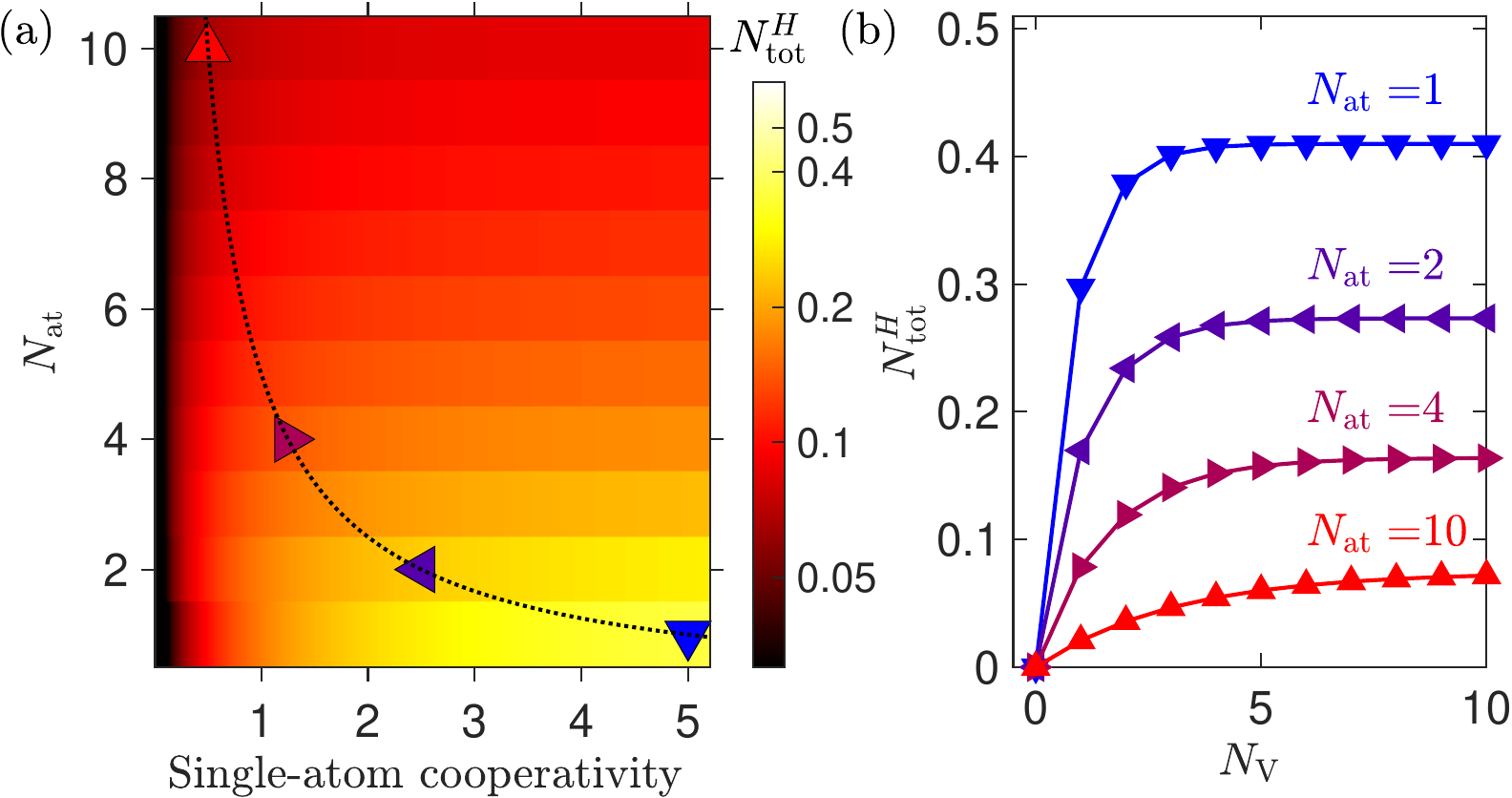}
\caption{(a) Total number of reflected $H$ photons $N_{\rm tot}^H$ in the limit $N_{\rm V}\to \infty$ depending on  the single-atom cooperativity $\gamma_{\rm 1D}/\gamma$ and the number of atoms $N_{\rm at}$.
(b) Dependence of $N_{\rm tot}^H$ on $N_{\rm V}$ for different values of $N_{\rm at}$. For each value of $N_{\rm at}$, 
we tune the decay $\gamma$ to keep constant the collective cooperativity, 
$C_{N_{\rm at}}=\gamma_{\rm 1D}(N_{\rm at}+1)/(2\gamma)=5$. Corresponding values of single-atom cooperativity $C_1=\gamma_{\rm 1D}/\gamma$ are shown in (a).
The calculation has been performed for $\omega=\omega_{0}$ and $B=0$.
}
\label{fig:3}
\end{figure}


We have performed a detailed simulation of the photon reflection process. 
We calculate the average projection of the total spin $\langle S_z\rangle$ and  the total number of reflected $H$ photons  $N_{\rm tot}^H(N_V)\equiv \sum_{j=1}^{N_V} R_{V\to H}^{(j)}$ depending on the number of incident $V$-polarized photons $N_V$. Here, the reflection coefficient $R_{V\to H}^{(j)}$ is the probability to reflect the $j$-th incident $V$-polarized photon with an $H$-polarization. 
Figure~\ref{fig:3} presents  the  total number of reflected $H$ photons $N_{\rm tot}^H$  given by 
\begin{align}\label{eq:N}
N_{\rm tot}^H(N_V)&=N_{\rm tot}^H(\infty)(1-|\chi|^{2N_V})\:,\\
N_{\rm tot}^H(\infty)&=\frac1{N_{\rm at}+1}\frac{N_{\rm at}C_{N_{\rm at}}}{N_{\rm at}C_{N_{\rm at}}+N_{\rm at}+1}\nonumber 
,\end{align}
\begin{figure}[t]
\centering\includegraphics[width=0.48\textwidth]{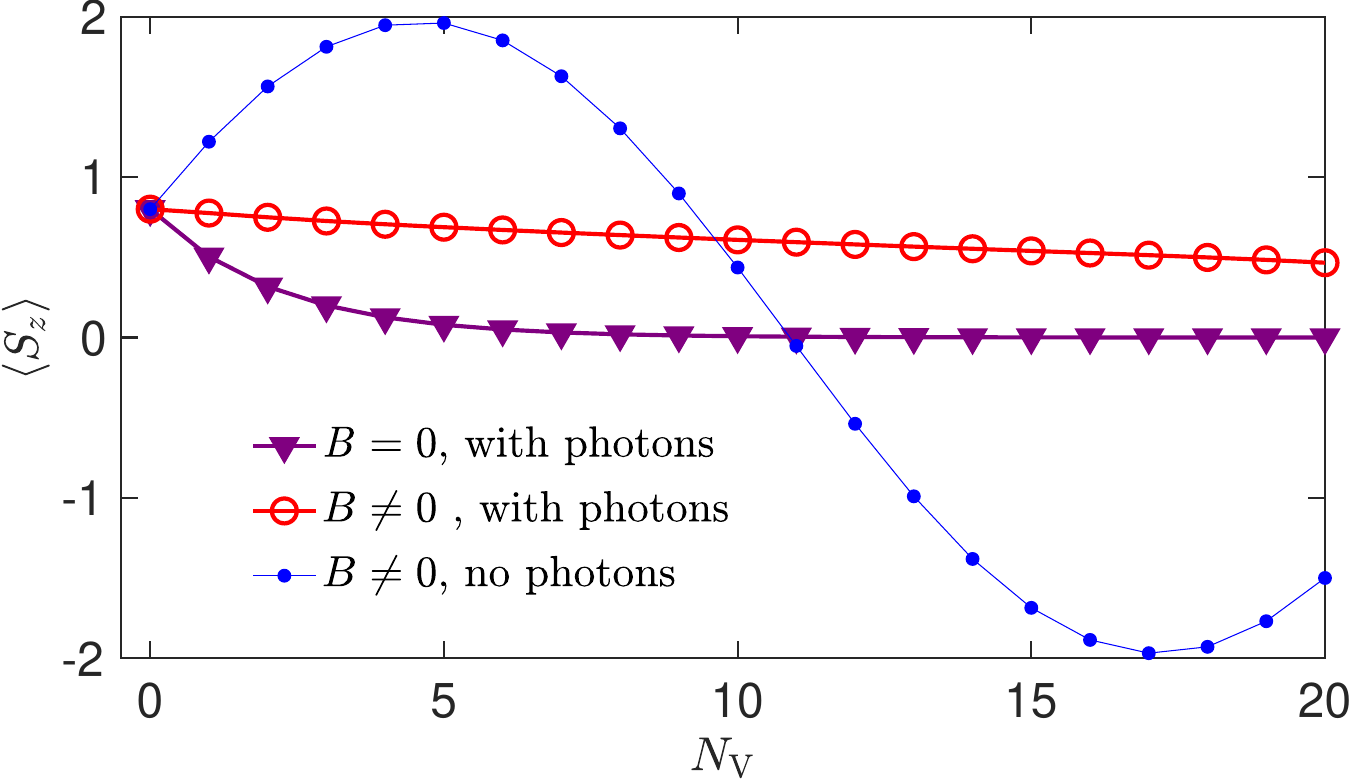}
\caption{ 
Average spin projection $\langle S_z\rangle$ calculated with and without applied external magnetic field $\bm B$ and incoming $V$ photons.
Calculation has been performed for the magnetic field strength parameter $\varphi=0.25$, $N_{\rm at}=4$ and $C_{N_{\rm at}}=5$.
}\label{fig:scanwh}
\end{figure}
where $\chi=1+\rmi\alpha$ (see  Supplemental Materials for the derivation).   In particular,  Fig.~\ref{fig:3}(a)  shows how  $N_{\rm tot}^H$ in the limit $N_{\rm V}\to \infty$  depends on the  number of atoms $N_{\rm at}$  and on the single-atom cooperativity $C_1$. One could expect  that this value would depend only on the  collective cooperativity $C_{N_{\rm at}}=N_{\rm at}C_1$.
This would mean that the same {outcome} could be achieved for a large $N_{\rm at}$ with low $C_1$ and a single atom with high $C_1$, provided that $C_{N_{\rm at}}$ stays the  same. However, this is not the case. The value of $N_{\rm tot}^H$ monotonously decreases with the number of atoms $N_{\rm at}$  along the line of constant  $C_{N_{\rm at}}$ (dotted curve in Fig.~\ref{fig:3}a). The same effect is illustrated in Fig.~\ref{fig:3}b that shows $N_{\rm at}$ versus $N_V$ at the  four points of constant collective cooperativity, indicated by triangles in Fig.~\ref{fig:3}a.  Not only does the limiting value of $N_{\rm tot}^H(N_V\to \infty)$ become smaller for larger $N_{\rm at}$, but also the   value of $N_V$ required to approach this limit increases. In other words, if  the $V$ photons arrive periodically in time, the dynamics of $N_{\rm tot}^H$ is slowed down.  The slowdown  is somewhat counterintuitive, since the photons are coupled only to collective Dicke states of the array that are typically  associated with the speedup of the atom-photon interaction. Here, however, the larger $N_{\rm at}$ and the stronger the interaction with the Dicke state, the slower  the dynamics. 
This happens because the relevant photon reflection process is quenched by the collectively enhanced spontaneous decay rate, see the increase of the denominator of
Eq.~\eqref{eq:alpha} with $N_{\rm at}$. It is this  collective slowdown of the spin evolution that we interpret as a Zeno effect responsible for  the suppression of the  single-photon nonlinearity  for large $N_{\rm at}$.

The QZE interpretation can be corroborated  by analyzing the dependence of the  $\langle S_z\rangle$ spin projection  on $N_V$, given by $S_z(N_V)=S_z(0)\Re (\chi^{N_V})$ or $S_z(N_V)=S_z(0)\Re \e^{\rmi\alpha N_V}$ for $\alpha\ll 1$. Thus,  after a large number of incident photons  the spin returns to the stationary value along $x$ axis.
The corresponding dynamics is shown by filled triangles in Fig.~\ref{fig:scanwh}. 
 It is  instructive  to study the effect of an external  static magnetic field along the $y$ direction.
Such a field leads to the rotation of the collective spin in the $x-z$ plane [see Fig.~\ref{fig:protocol}(c)], which competes with the Zeno effect.  To describe this  we introduce  a total spin rotation  $\psi'=\exp(\rmi \varphi S_y)\psi$ to the wavefunction between the photon scattering
events. Here, 
$\varphi$ is the spin rotation angle during the time between the two incident $V$ photons. This simplified description assumes that the Zeeman splitting is much smaller than the resonance linewidth and optical selection rules are not modified.
The calculated spin  dynamics is  shown in Fig.~\ref{fig:scanwh} by the circles.
Small filled circles show the   free spin rotation  induced by the magnetic field without any  incoming photons. When the $V$-polarized photons are sent upon the system (red open circles) the oscillations are replaced by the slow decay of the spin toward the equilibrium state.  This further supports our interpretation  in terms of the  quantum  Zeno effect.

\begin{acknowledgements}
{We gratefully acknowledge stimulating discussions with Jeff Thompson, Janet Zhong, Nikita Leppenen, Alexander Poshakinskiy, and Ephraim Shahmoon. 
ANP acknowledges support by the Center for New Scientists at the Weizmann Institute of Science. BD acknowledges support from the Israel Science Foundation, Minerva foundation, and the US-Israel Binational Science Foundation. BD is the Dan Lebas and Roth Sonnewend Professorial Chair of Physics.}
\end{acknowledgements}

%

\setcounter{figure}{0}
\setcounter{section}{0}
\setcounter{equation}{0}
\renewcommand{\thefigure}{S\arabic{figure}}
\renewcommand{\thesection}{S\Roman{section}}
\renewcommand{\thesection}{S\arabic{section}}
\renewcommand{\theequation}{S\arabic{equation}}

\newpage\clearpage
\section*{Supplementary materials}

\subsection{General approach for the scattering amplitude calculation}
Here we discuss in more detail the calculation of the single-photon scattering matrix for an atomic array. This can be done by a straightforward generalization of the  Green function approach for two-level atoms coupled to the waveguide, see  e.g.  Ref.~\cite{Kocabas2016} and  Appendix F of Ref.~\cite{sheremet2021waveguide}.

\begin{figure}[t]
\centering\includegraphics[width=0.2\textwidth]{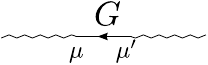}
\caption{ Diagram representing the reflection of a single photon from an array of atoms. The bold straight line represents the dressed Green's function of the array in the single-excited state, and the wavy lines are Green functions of the incoming and outgoing photons. Indices $\mu$, $\mu'$ label different single-excited states.}\label{fig:S1}
\end{figure}

The corresponding diagram representing the single-photon scattering is shown in Fig.~\ref{fig:S1}.
The scattered state $S\psi_0$ is given by
\begin{equation}\label{eq:S} 
S(\omega)\psi_0=\psi_0+\langle f|G(\omega)|i\rangle\psi_0\:.
\end{equation}
where $\psi_0$ is the state before scattering and $S$ is the scattering matrix. The second term in Eq.~\eqref{eq:S} describes photon interaction with the  single-excited states of the array with
$G$ being the corresponding Green function of these states. 
It is given by inverting the Hamiltonian for single-excited states,
\begin{equation} 
G(\omega)=\rmi \gamma_{\rm 1D}\sum\limits_{\mu\mu'} |\psi_1^{(\mu)}\rangle \left[\frac1{H_1-\hat 1\omega}\right]_{\mu\mu'}\langle\psi_1^{(\mu')}|\:.
\end{equation}
The states $|i\rangle$ and $|f\rangle$ are determined by the spatial distribution of the electric field and the polarization of the incoming and outgoing photons:
\begin{equation}\label{eq:if}
|i\rangle=\sum\limits_{n=1}^{N_{\rm at}}E^{(i)}_{n,\nu}\sigma_{n,\nu}^\dag\:,
|f\rangle=\sum\limits_{n=1}^{N_{\rm at}}E^{(f)}_{n,\nu}\sigma_{n,\nu}^\dag\:,    
\end{equation}
where $E^{(i)}_{n,\nu}$ is the electric field amplitude of the incoming ($i$) and outgoing ($f$) photon with polarization $\nu$ at the $n^{\textrm{th}}$ atom. We are interested in the case when all atoms are located in one point. In this case, the electric field at all the atoms is the same, $E^{(i)}_{n,\nu}=E^{(f)}_{n,\nu}=1$, and Eq.~\eqref{eq:if} simplifies to
\begin{equation}
|i\rangle=\sigma_{\nu}^{\rm tot,\dag}    ,\quad |f\rangle=\sigma_{\nu'}^{\rm tot,\dag}
\end{equation}
with $\sigma_{\nu}^{\rm tot,\dag}\equiv \sum_n \sigma_{n,\nu}^\dag$ and $\nu$ ($\nu'$) are the polarizations of incoming (outgoing) photons.
The result is 
\begin{equation}\label{eq:finalS}
\psi_{\rm scat,\nu\rightarrow\nu'}=[\delta_{\nu,\nu'}+\sigma^{\rm tot}_{\nu'} G(\omega) \sigma^{\rm tot,\dag}_{\nu}]   \psi_{\rm ground},
\end{equation}
This is equivalent to Eqs.~(3,4) in the main text.
The operator $\delta_{\nu,\nu'}+\sigma^{\rm tot}_{\nu'} G(\omega) \sigma^{\rm tot,\dag}_{\nu} $ in Eq.~\eqref{eq:finalS} is similar to the Kraus operator  in the quantum measurement theory \cite{Kraus1981,Bednorz_2012}.

In order to illustrate the general approach we   perform below the calculation explicitly for two particular cases of $N=1$ and $N=2$ atoms. 
\subsubsection*{Scattering on $N=1$ atom.}
We label the two possible ground states and the single-excited state of the atom as 
\begin{equation}
\psi_{\rm ground}^{(1)}=|+\rangle, \psi_{\rm ground}^{(2)}=|-\rangle, \psi_{1}^{(2)}=|0\rangle\:.
\end{equation}
We are interested in the photon interaction with the following symmetric superposition of the two
ground states
\begin{equation}\label{eq:psi_ground0}
\psi_{\rm ground}=\frac1{\sqrt{2}}(|+\rangle+|-\rangle)\:.\end{equation}
The effective Hamiltonian for single-excited states is given just by $H_1=\omega_0-\rmi \gamma_{\rm 1D}$.
If the atom is excited by a single $\sigma_+$--polarized photon,
there are two possible scattered states. 
The one where the scattered photon has conserved polarization is
\begin{equation}
\psi_{\rm scat,+\rightarrow +}=\psi_{\rm ground} +\rmi\sigma_{\rm tot,+}\frac{\rmi \gamma_{\rm 1D}}{H_1-\omega}\sigma_{\rm tot,+}^\dag 
\psi_{\rm ground}
\end{equation}
and the one where the scattered photon  is $\sigma_-$ is
\begin{equation}
\psi_{\rm scat,+\rightarrow -}=\sigma_{\rm tot,-}\frac{\rmi \gamma_{\rm 1D}}{H_1-\omega}\sigma_{\rm tot,+}^\dag\psi_{\rm ground} \:.
\end{equation}
Explicit calculation results in 
\begin{align}\label{eq:psiscatN1}
\psi_{\rm scat,+\rightarrow +}&=\frac{1}{\sqrt{2}}
\left(\frac{\omega-\omega_0}{\omega-\omega_0+\rmi\gamma_{\rm 1D}+\rmi\gamma}\psi_{\rm ground}^{(1)}+\psi_{\rm ground}^{(2)}\right)\:,\nonumber\\
\psi_{\rm scat,+\rightarrow -}&=-\frac{1}{\sqrt{2}}
\frac{\rmi\gamma_{\rm 1D}}{\omega-\omega_0+\rmi\gamma_{\rm 1D}+\rmi\gamma}\psi_{\rm ground}^{(2)}\:.
\end{align}
It can be easily checked that the scattering is unitary for $\gamma=0$, that is
$\sum_{\nu'}\langle \psi_{\rm scat,+\rightarrow\nu'}|\psi_{\rm scat,+\rightarrow\nu'}\rangle=1$. This also agrees with the results in Ref.~\cite{Zhong2023}.
\subsubsection*{Scattering on $N=2$ atoms.}
This case is a bit less trivial since there are  four possible ground states of the array, given by
\begin{equation}
\psi_{\rm ground}^{(1,2,3,4)}=|+_1,+_2\rangle, |+_1,-_2\rangle, |-_1,+_2\rangle, |-_1,-_2\rangle\:.
\end{equation}
The considered ground state is a symmetric combination 
\begin{equation}\label{eq:psi_ground1}
\psi_{\rm ground}=\frac{1}{2}\sum\limits_{j=1}^4\psi^{(j)}_{\rm vac}\equiv
\frac1{2}(|+_1\rangle+|-_1\rangle)(|+_2\rangle+|-_2\rangle)\:,
\end{equation}
and the single-excited states are given by 
\begin{align}\label{eq:psi1N2}
\psi_{1}^{(1)}=|+_1,0_2\rangle,\quad  &\psi_{1}^{(2)}=|-_1,0_2\rangle\:,\\ \psi_{1}^{(3)}=|0_1,+_2\rangle,\quad  &\psi_{1}^{(4)}=|0_1,-_2\rangle.\nonumber
\end{align}
The effective Hamiltonian for single-excited states $H_1$ is given by projecting the full Hamiltonian onto the basis Eq.~\eqref{eq:psi1N2} and explicitly reads
\begin{equation}
H_1=\omega_0-\rmi\gamma_{\rm 1D}\begin{pmatrix} 1&0&\frac{1}{2}&0\\0&1&0&\frac{1}{2}\\
\frac{1}{2}&0&1&0\\
0&\frac{1}{2}&0&1
\end{pmatrix}\:.
\end{equation}
The array is initially in the state Eq.~\eqref{eq:psi_ground1} and is excited by one $\sigma_+$--polarized photon. Similarly to Eqs.~\eqref{eq:psiscatN1}, we find the scattered state of atomic array with the conserved photon polarization
\begin{multline}
\psi_{\rm scat,+\rightarrow +}=\psi_{\rm ground} 
\\-\frac{\rmi\gamma_{\rm 1D}}{\omega-\omega_0+\frac{3\rmi}{2}\gamma_{\rm 1D}+
\rmi\gamma}(2\psi_{\rm ground}^{(1)}+\psi_{\rm ground}^{(2)}+\psi_{\rm ground}^{(3)})\:.
\end{multline}
The amplitude of the process where the scattered photon  has changed polarization is
\begin{equation}
\psi_{\rm scat,+\rightarrow -}=-\frac{\rmi\gamma_{\rm 1D}}{\omega-\omega_0+\frac{3\rmi}{2}\gamma_{\rm 1D}+
\rmi\gamma}\sum_{j=2}^4\psi_{\rm ground}^{(j)}\:.
\end{equation}
It is again straightforward to check that the scattering is unitary when $\gamma=0$, 
$\sum_{\nu'}\langle \psi_{\rm scat,+\rightarrow\nu'}|\psi_{\rm scat,+\rightarrow\nu'}\rangle=1$.
An important difference from the single-atom case, Eqs.~\eqref{eq:psiscatN1}, is that the scattering amplitude
has a pole at the complex frequency
$\omega_0-3\rmi\gamma_{\rm 1D}/2$ with the imaginary part being larger by the absolute value than the value for a single atom. This reflects the superradiant enhancement of the radiative decay rate. The general expression for the decay rate of an $N$-atom array is 
$(N+1)\gamma_{\rm 1D}/2$.

\subsection{Array in a cavity}
In this section, we show  that the reflection coefficient for an array of two-level atoms in a cavity obtained by the approach from the previous section is equivalent to the one given by the Airy formula for the multiple reflections. This provides an independent verification of our method.

We consider a two-level atom array, so that the Hamiltonian becomes
\begin{equation} \label{eq:H1b}
\hat H={\omega_0\sum\limits_{n=1}^{N_{\rm at}}|0,n \rangle \langle 0,n|}-\rmi\frac{\gamma_{\rm 1D}}{2}\sigma^{{\rm tot},\dag}\sigma^{{\rm tot}}\:.
\end{equation}
Comparing with Eq.~(2) in the main text, we see that the polarization index $\nu$ has been suppressed. The reflection coefficient is  calculated according to Eq.~(4) in the main text [supplementary Eq.~\eqref{eq:S}]:
\begin{equation} \label{eq:r1b}
r=1+\frac{\rmi \gamma_{\rm 1D}}{H_1-\omega}\:.
\end{equation}
The Hamiltonian $H_1$ of single-excited states is given by $[H_1]_{mn}=\omega_0\delta_{mn}-\rmi N\gamma_{\rm 1D}/2$\:.
As a result, the reflection coefficient assumes the form
\begin{equation} \label{eq:r1b2}
r=
\frac{\omega_0+\rmi N\gamma_{\rm 1D}-\omega}{\omega_0-\rmi N\gamma_{\rm 1D}-\omega}\:.
\end{equation}
Now we will obtain the same answer using a different approach. We start from the expressions  for the reflection and transmission amplitudes of an array of two-level atoms, coupled to a waveguide without any cavity~\cite{sheremet2021waveguide}\:,
\begin{equation} \label{eq:r0b}
r_0=\frac{\rmi N\gamma_{\rm 1D}^{(0)}/2}{\omega_0-\omega-\rmi N\gamma_{\rm 1D}^{(0)}/2}, \quad t_0=r_0+1\:,
\end{equation}
where $\gamma_{\rm 1D}^{(0)}/2$ is the spontaneous emission rate into the waveguide. Next, we put such an array at an integer wavelength spacing from a left side of a perfect mirror. The reflection coefficient from the combined system ``array+mirror'' can be calculated by summing the amplitude process where light has been reflected once from the array and those where light has been transmitted through the array, bounced between the array and the mirror, and then transmitted back, see e.g.,  Ref.~\cite{semina}. The sum of these contributions is
\begin{equation} \label{eq:r0c}
r_1=r_0+\frac{t_0^2}{1-r_0}=\frac{\omega_0+\rmi N\gamma_{\rm 1D}^{(0)}-\omega}{\omega_0-\rmi N\gamma_{\rm 1D}^{(0)}-\omega}\:.
\end{equation}
We see, that, compared to Eq.~\eqref{eq:r0b}, the spontaneous decay rate has increased twice in the vicinity of the mirror. The absolute value of the reflection coefficient
is now unity because the mirror is fully reflecting.
Next, we add one more mirror to the left of the array at the integer wavelength spacing. This second mirror is characterized by the reflection coefficient $\rho$ and the transmission coefficient $\tau^2=1-\rho^2$. 
The reflection coefficient is found similar to Eq.~\eqref{eq:r0c}
\begin{equation}
\widetilde r=-\rho+\tau^2 \frac{r_1}{1-\rho r_1}\:,
\end{equation}
and reads
\begin{equation}
\widetilde r=\frac{\omega_0+\rmi N\gamma_{\rm 1D}-\omega}{\omega_0-\rmi N\gamma_{\rm 1D}-\omega},\quad 
\gamma_{\rm 1D}=2\gamma_{\rm 1D}^{(0)}\frac{1+\rho}{1-\rho}\:.
\end{equation}
This answer is equivalent to Eq.~\eqref{eq:r1b2} if we interpret  $\gamma_{\rm 1D}$ as the spontaneous decay rate modified by the presence of a cavity.
\subsection{Single-photon scattering from an array}
Here we provide more detail on the calculation of the single photon scattering on the array with $N_{\rm at}$ atoms located at the same point (or coupled to the same photonic mode of the cavity).
{In order to evaluate the operators $\sigma^{\rm tot}_{\nu'} G(\omega) \sigma^{\rm tot,\dag}_{\nu}$ entering Eq.~\eqref{eq:final} we first calculate the following operator products:
\begin{align}
&\sigma^{j}_{+} \sigma^{j',\dag}_{-}=\delta_{jj'}(s_x^{(j)}+\rmi s_y^{(j)})\:,
&\sigma^{j}_{-} \sigma^{j',\dag}_{+}=\delta_{jj'}(s_x^{(j)}-\rmi s_y^{(j)})\:,\nonumber\\
&\sigma^{j}_{H} \sigma^{j',\dag}_{V}=\delta_{jj'} (s_z^{(j)}-\rmi s_y^{(j)})\:,
&\sigma^{j}_{V} \sigma^{j',\dag}_{H}=\delta_{jj'} (s_z^{(j)}+\rmi s_y^{(j)})\:,\nonumber\\
&\sigma^{j}_{H} \sigma^{j',\dag}_{H}=\delta_{jj'} (s+s_x^{(j)})\:,
&\sigma^{j}_{V} \sigma^{j',\dag}_{V}=\delta_{jj'} (\tfrac{1}{2}-s_x^{(j)})\:,\nonumber\\
&\sigma^{j}_{+} \sigma^{j',\dag}_{+}=\delta_{jj'} (\tfrac{1}{2}+s_z^{(j)})\:,
&\sigma^{j}_{-} \sigma^{j',\dag}_{-}=\delta_{jj'} (\tfrac{1}{2}-s_z^{(j)})\:.\label{eq:selrules1}
\end{align}
Here, $s_x^{(j)}$, $s_y^{(j)}$ and  $s_z^{(j)}$ are the psedospin-1/2 operators acting on the two ground states of the atom $j$. Equations~\eqref{eq:selrules1} can be straightforwardly checked by evaluating the left-hand side for an arbitrary ground state of the ensemble.

Next, we analyze the single-excited states excited by photons. In the considered case
 the atom-photon interaction involves only the sum of the excitation operators, $\sigma_\nu^{{\rm tot}}=\sum_{n=1}^{N_{\rm at}}\sigma_{\nu,n}$. Hence, the only relevant single-excited  states are the symmetric  combinations such as 
 \begin{multline}\label{eq:psi1exc}
\psi_1^{(1)}=\frac1{\sqrt{N_{\rm at}}}[|0_1+_2+_3\ldots +_{N_{\rm at}}\rangle\\+
|+_10_2+_3\ldots +_{N_{\rm at}}\rangle\\+\ldots+|+_1+_2+_3\ldots 0_{N_{\rm at}}\rangle]\:,
\end{multline}
\begin{multline}\label{eq:psi2exc}
\psi_1^{(2)}=\frac1{\sqrt{N_{\rm at}(N_{\rm at}-1)}}[|0_1-_2+_3\ldots +_{N_{\rm at}}\rangle\\+
|-_10_2+_3\ldots +_{N_{\rm at}}\rangle+\ldots]\:,
\end{multline}
and so on. They can be distinguished by the number of atoms in the state  ```$-$''  in each of the term of the sum. This number runs from 0, as in Eq.~\eqref{eq:psi1exc} to $N_{\rm at}-1$ yielding 
 $N_{\rm at}$ such single-excited states in total. All these states are the eigenstates of the Hamiltonian Eq.~\eqref{eq:H1} in the main text with the eigenvalue given by  ($\gamma=0$)
\begin{equation}\label{eq:psi1excH}
\langle \psi_1^{(\mu)}|
H|\psi_1^{(\mu)}\rangle= -\rmi \gamma_{\rm 1D}\frac{N_{\rm at}+1}{2},\quad \mu=1\ldots N_{\rm at}.
\end{equation}
 Equation \eqref{eq:psi1excH} can be most easily checked for the state Eq.~\eqref{eq:psi1exc}. Specifically,
\begin{multline}\label{eq:psi1excH2}
{\langle\psi_1^{(1)} |\sigma_{\rm tot,+}^\dag \sigma_{\rm tot,+}  |\psi_1^{(1)}\rangle}=
  \sum\limits_{j,j'=1}^{N_{\rm at}}\langle\psi_1^{(1)}|\sigma^{j}_{+} \sigma^{j',\dag}_{+}|\psi_1^{(1)}\rangle\\=
\sum\limits_{j,j'=1}^{N_{\rm at}}\frac1{N_{\rm at}}
=N_{\rm at}.
\end{multline}
\begin{multline}\label{eq:psi1excH2b}
\langle\psi_1^{(1)} |\sigma_{\rm tot,-}^\dag \sigma_{\rm tot,-}|\psi_1^{(1)}\rangle=
  \sum\limits_{j,j'=1}^{N_{\rm at}}\langle\psi_1^{(1)}|\sigma^{j}_{-} \sigma^{j',\dag}_{-}|\psi_1^{(1)}\rangle\\
  =\sum\limits_{j,j'=1}^{N_{\rm at}} \frac{\delta_{j,j'}}{N_{\rm at}}=N_{\rm at}.
\end{multline}
Since $H=-\rmi \gamma_{\rm 1D}\sum_{\nu=\pm}\sigma_{\rm tot,\nu}^\dag \sigma_{\rm tot,\nu}/2$, a sum of Eq.~\eqref{eq:psi1excH2} and Eq.~\eqref{eq:psi1excH2b} yields Eq.~\eqref{eq:psi1excH}. 
Because of the  rotation symmetry in the space spanned by the ground states of the ensemble,  Eq.~\eqref{eq:psi1excH} is valid also for the other values of $\mu$.
 
Using Eqs.~\eqref{eq:selrules1}  and Eq.~\eqref{eq:psi1excH} we readily evaluate the operators 
$\sigma^{\rm tot}_{\nu'} G(\omega) \sigma^{\rm tot,\dag}_{\nu}$ 
describing  single photon scattering for all  incident and scattered photon polarizations}
\begin{align}
&\sigma^{\rm tot}_{+} G\sigma^{{\rm tot},\dag}_{-}=\rmi\alpha (S_x+\rmi S_y),
&\sigma^{\rm tot}_{-} G\sigma^{{\rm tot},\dag}_{+}=\rmi\alpha (S_x-\rmi S_y),\nonumber\\
&\sigma^{{\rm tot}}_{H} G\sigma^{{\rm tot},\dag}_{V}=\rmi\alpha (S_z-\rmi S_y),
&\sigma^{{\rm tot}}_{V} G\sigma^{{\rm tot},\dag}_{H}=\rmi\alpha (S_z+\rmi S_y),\nonumber\\
&\sigma^{{\rm tot}}_{H} G\sigma^{{\rm tot},\dag}_{H}=\rmi\alpha (S+S_x),
&\sigma^{{\rm tot}}_{V} G\sigma^{{\rm tot},\dag}_{V}=\rmi\alpha (S-S_x)\nonumber\\
&\sigma^{{\rm tot}}_{+} G\sigma^{{\rm tot},\dag}_{+}=\rmi\alpha (S+S_z),
&\sigma^{{\rm tot}}_{-} G\sigma^{{\rm tot},\dag}_{-}=\rmi\alpha (S-S_z)\:.\label{eq:selrules}
\end{align}
Here 
\begin{equation}\label{eq:alpha2}
 \alpha(\omega) =-\frac{\gamma_{\rm 1D}}{\omega-\omega_0+\frac{\rmi}{2}\gamma_{\rm 1D}(N_{\rm at}+1)}\:
\end{equation}
is the effective polarizability of the array. 
For nonzero decay rate $\gamma$ we can just replace $\omega_{0}$ by $\omega_{0}-\rmi \gamma$ in Eq.~\eqref{eq:alpha2}.
Equation~\eqref{eq:selrules} generalizes Eq.~(6) from the main text. Next, we use the basis of the states 
$|S,S_z\rangle$ with  the total (pseudo)spin $S$ and spin projection $S_z$ to describe the ground state of the array:
\begin{equation}
\psi_{\rm ground}=|S=N_{\rm at}/2, S_x=N_{\rm at}/2\rangle\:.
\end{equation}
The same  state can be equivalently presented as 
\begin{equation}\label{eq:zx}
|S_x=S/2\rangle=\frac1{2^S}\sum\limits_{S_z=-S}^S  (-1)^{S_z}\sqrt{C_{2S}^{2S_z}} |S_z\rangle\:,
\end{equation}
see, e.g., \S 4.14 in Ref.~\cite{VMKen}.
We remind that 
\begin{multline}
\langle S,S_z|S_{z,+} |S,S-1\rangle=\langle S,S_z-1|J_{z,-} |S,S_z\rangle\\= \sqrt{(S+S_z)(S-S_z+1)}\:.
\end{multline} Using this expression and Eq.~\eqref{eq:S} we find that after the scattering of the first $\sigma_+$-polarized photon with the change of polarization, the state of the array becomes 
\begin{multline}\label{eq:zx1}
|1\rangle\equiv \langle S_x|\sigma^{{\rm tot}}_{-} G\sigma^{{\rm tot},\dag}_{+}|S_x=N_{\rm at}/2\rangle\\=\frac{\rmi\alpha}{2} \left[
N_{\rm at} |M_x=N_{\rm at}/2\rangle+\sqrt{N_{\rm at}} |M_x=N_{\rm at}/2-1\rangle
\right]\:.
\end{multline}
The total scattering amplitude can be found as
$\langle 1|1\rangle$
and reads
\begin{equation}
R_1=|\alpha^2|\frac{N_{\rm at}(N_{\rm at}+1)}{4}\:.
\end{equation}
Clearly, for large $N_{\rm at}$ the scattering probability tends to unity. 
We now normalize this wavefunction for $\gamma=0 $assuming that the photon $\sigma_+$ has been scattered indeed. The result is 
\begin{equation}\label{eq:step1b}
|1'\rangle=
\sqrt{\tfrac{N_{\rm at}}{N_{\rm at}+1}} \bigl|S_x=\tfrac{N_{\rm at}}{2}\bigr\rangle+\sqrt{\tfrac{1}{N+1}} \bigl|S_x=\tfrac{N_{\rm at}}{2}-1\bigr\rangle
\end{equation}
This state carries a total spin 
\begin{align}\label{eq:total_spin}
&\bm S=\frac{N_{\rm at}}{2}\bm e_x-\delta S_x\bm e_x+\delta S_z\bm e_z, \\ &\delta S_x=\frac1{N_{\rm at}+1},\quad \delta S_z=\frac{N_{\rm at}}{N_{\rm at}+1}\:.\nonumber
\end{align}
estimated as
\begin{equation}
\delta S_x=S(1-\cos \varphi)\approx \frac1{N_{\rm at}},\quad \varphi=\frac{2}{N_{\rm at}} \:,
\end{equation}
where $\varphi$ is the angle of total spin rotation around $y$  from $x$ towards $z$ after $L$ photon has flipped.

Next, after scattering a $V$ photon, the state becomes 
\begin{equation}\label{eq:step2}
|2\rangle\equiv   \sigma^{{\rm tot}}_{H} G\sigma^{{\rm tot},\dag}_{V}|1\rangle=\rmi\alpha
\sqrt{\tfrac{N_{\rm at}}{N_{\rm at}+1}}  |S_x=\tfrac{N_{\rm at}}{2}-1\rangle\:.
\end{equation}
The scattering amplitude is 
\begin{equation}
R^{(1)}_{V\rightarrow H}=|\alpha^2|\frac{N_{\rm at}}{N_{\rm at}+1}=\frac{4}{(N_{\rm at}+1)^2}\:.
\end{equation}

\subsection{Scattering of multiple $V$ photons}
In this section we derive Eq.~(6) for the spin dynamics, given in the main text.
Equation~\eqref{eq:step1b} shows that it is sufficient to describe the wavefunction in the basis of just two spin states, $S_x=N_{\rm at}/2$ and  $S_x=N_{\rm at}/2-1$. The density matrix, corresponding to the state Eq.~\eqref{eq:step1b} at the initial stage is given by
\begin{equation}
\rho_0=\begin{pmatrix}\frac{N_{\rm at}}{N_{\rm at}+1}&\frac{\sqrt{N_{\rm at}}}{N_{\rm at}+1}\\\frac{\sqrt{N_{\rm at}}}{N_{\rm at}+1}&\frac{1}{N_{\rm at}+1}\end{pmatrix}.
\end{equation}
The spin operators in the reduced basis are
\begin{equation}
S_z=\frac{\sqrt{N_{\rm at}}}{2}\begin{pmatrix}0&1\\1&0\end{pmatrix},\quad 
S_x=\begin{pmatrix}\frac{N_{\rm at}}{2}&0\\0&\frac{N_{\rm at}}{2}-1\end{pmatrix}\:.
\end{equation}
The wavefunction at each step is multiplied by the Kraus operator
\begin{multline}
K=1+\sigma^{(\rm tot)}_{V} G\sigma^{(\rm tot),\dag}_{V}\\=\chi\bigl|S_x=\tfrac{N_{\rm at}}{2}-1\bigr\rangle\bigl\langle S_x=\tfrac{N_{\rm at}}{2}-1\bigr|+\bigl|S_x=\tfrac{N_{\rm at}}{2}\bigr\rangle\bigl\langle S_x=\tfrac{N_{\rm at}}{2}\bigr|\:,
\end{multline}
where $\chi=1+\rmi \alpha$.
This leads to the following evolution of the density matrix
\begin{gather}
\rho(N_V+1)=\begin{pmatrix}
1-|\chi^2|\rho_{22}(N_V)&\rho_{1,2}(N_V)\rmi \alpha\\
\rho_{1,2}(N_V)\rmi \alpha^*&\rho_{2,2}(N_V)|\chi^2|
\end{pmatrix}\:,
\end{gather}
and  the following spin dynamics
\begin{equation}
S_z(N_V)=\frac{N_{\rm at}}{N_{\rm at}+1}\Re\chi ^{N_V}\:.
\end{equation}
After $N_V$ photons have been scattered the reflection coefficient is given by
\begin{equation}
R_{V\to H}^{(N_V)}=\frac1{N_{\rm at}+1}\frac{N_{\rm at}}{N_{\rm at}+2\gamma}\left[1-|\chi|^{2(N_V-1)}\right]\:.
\end{equation}
Summing up over all the values of $N_V$ we obtain the second of Eqs.~\ref{eq:N} from the main text.

\subsection{Role of magnetic field}
\begin{figure}[t]
\centering\includegraphics[width=0.48\textwidth]{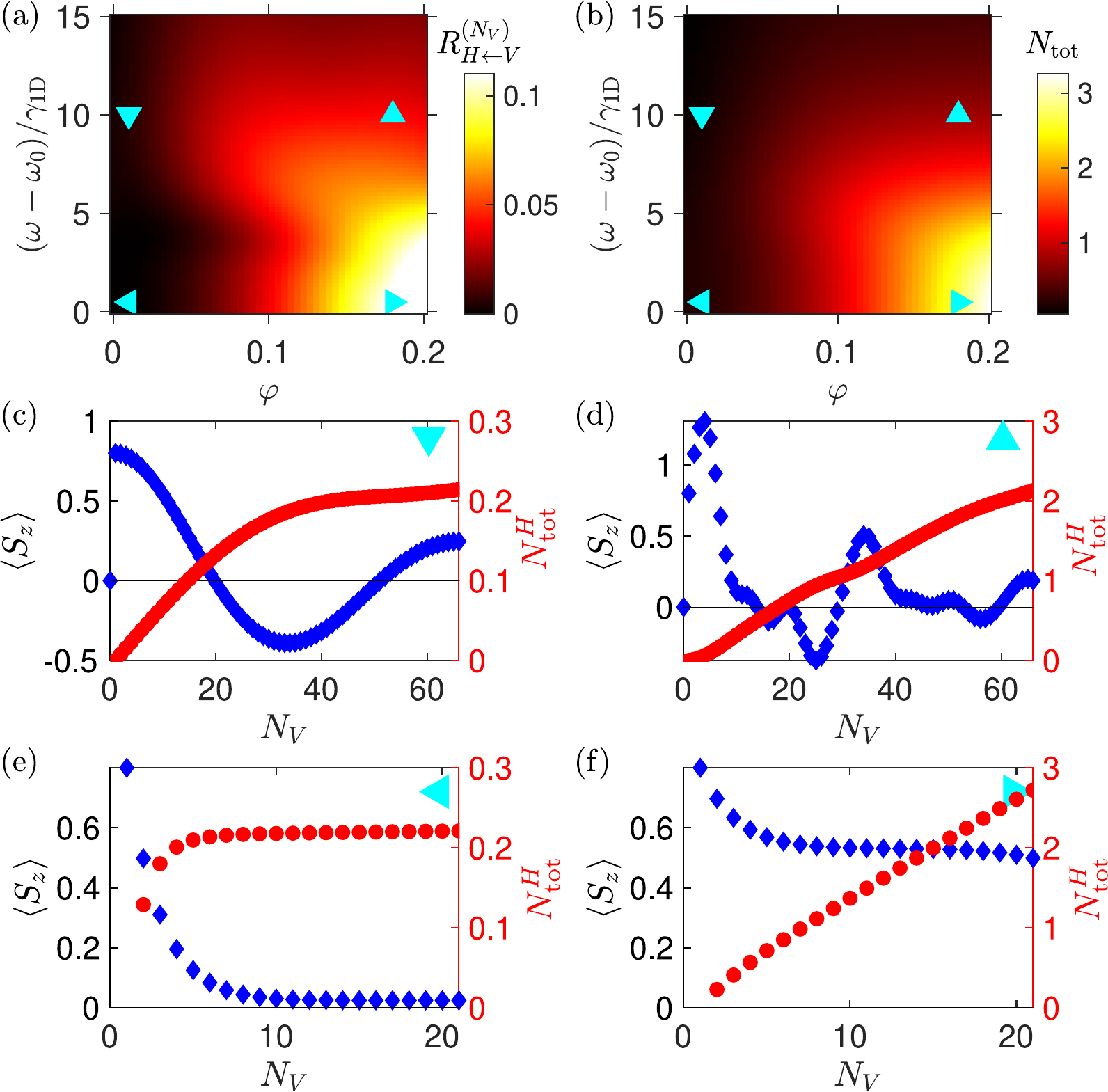}
\caption{ Color maps of the
reflection coefficient $R_{V\to H}$  (a) and total number of photons reflected $N_{\rm tot}^H$(b) after $N_V=20$ $V$ photons have been sent,   depending on the photon frequency $\omega$ and the phase $\varphi$ induced by the magnetic field along $y$. Calculated for $N_{\rm at}=4$
atoms.  (c--f) Dynamics of the total spin $\langle S_z\rangle$ (diamonds, left $y$-scale) and
 the total number of reflected photons $N_{\rm tot}^H$  (circles, right $y$-scale) depending on the number of $N_V$ photons sent. The calculations have been performed for  $(\omega-\omega_0)/\gamma$ and $\varphi$ equal to $(10,0.01)$; $(10,0.18)$; $(0.5,0.01)$; $(0.5,0.18)$, respectively (see also   triangles in (a)).
}\label{fig:scanwh1}
\end{figure}
Here we discuss in more detail the effect of   non-zero  external magnetic field, complementing 
Fig.~\ref{fig:scanwh} in the main text.  The results of simulation are shown in Fig.~\ref{fig:scanwh1}. Panels (a) and (b) present the color maps of the photon reflection coefficient depending on the dimensionless magnetic field strength $\varphi$ and the  detuning from the resonance, $(\omega-\omega_0)/\gamma_{\rm 1D}$. Figs.~\ref{fig:scanwh1}(c--f) show the dynamics of the spin and photon reflection for four characteristic values of the field strength and the detuning, indicated by triangle symbols in Fig.~\ref{fig:scanwh1}(a,b). 
When the detuning is small and the magnetic field is absent, the spin projection $S_z$  monotonously decays to zero and the total number of photons saturates at a constant value, see  Fig.~\ref{fig:scanwh1}(e). 
This  agrees with the results in Fig.~\ref{fig:3} in the main text, calculated for zero magnetic field. 
Larger detuning leads to a  slower decay and oscillatory behavior of $S_z$, instead of a  monotonous decay at $\omega=\omega_0$, as shown in   Fig.~\ref{fig:scanwh1}(c). 
Introduction of the magnetic field significantly modifies the spin dynamics. This can be seen in Fig.~\ref{fig:scanwh1}(f), corresponding to larger magnetic field and small detuning. Instead of decaying with $N_V$, as in 
Figs.~\ref{fig:scanwh1}(c,e), the spin projection freezes at the value $\langle S_z\rangle\approx 0.6$. At the  same time, the total number of reflected photons $N_{\rm tot}^H$ does not saturate but keeps linearly increasing  with $N_V$. This increase is also manifested as the bright spots in the bottom-right corners of the color maps Fig.~\ref{fig:scanwh1}(a,b) for $R_{V\to H}$ (a)  and $N_{\rm tot}^H$ (b). The  reflection coefficient does not  decay to zero  because of the magnetic field-induced rotation, that keeps the spin away from the dark state along the $x$ axis. The  slowdown of the spin projection dynamics in Fig.~\ref{fig:scanwh1}(f) supports our interpretation of the photon scattering  in terms of a  quantum  Zeno effect.

If the magnetic field is kept nonzero but the detuning is increased,  the spin projection demonstrates 
a complex oscillatory behaviour, as  shown in Fig.~\ref{fig:scanwh1}(c). 
 Such oscillations result from  the beatings between different collective Zeeman levels, split by the magnetic field. The total number of reflected photons $N_{\rm tot}^H$ still does not saturate with $N_V$ but is  smaller than in  Fig.~\ref{fig:scanwh1}(d) because of the weaker atom-photon interaction at larger detuning.

\end{document}